\documentstyle[epsf,osa,manuscript]{revtex}  
\tightenlines

\begin{document}                

\preprint{\it Journal of the Optical Society of America\/ \bf B 12\rm, 
821--828 (1995).}

\title{Spin Correlated Interferometry on Beam Splitters:\\ 
Preselection of Spin Correlated Photons} 

\author{Mladen Pavi\v ci\'c}

\address{Institut f\"ur Theoretische Physik, 
TU Berlin, Hardenbergstra\ss e 36, D--10623 Berlin 12, Germany\\
Atominstitut der \"Osterreichischen Universit\"aten, 
Sch\"uttelstra\ss e 115, A--1020 Wien, Austria\\
and Department of Mathematics, University of Zagreb, 
GF, Ka\v ci\'ceva 26, HR--41001 Zagreb, 
Croatia$^*$\footnotetext[1]{\ Permanent address. 
E-mail: mpavicic@faust.irb.hr; Web-page: http://m3k.grad.hr/pavicic}}

\maketitle

\begin{abstract}
A nonclassical feature of the fourth--order interference at a beam 
splitter, that genuine photon spin singlets are emitted in predetermined 
directions even when incident photons are unpolarized, has been used 
in a proposal for an experiment that imposes quantum spin correlation 
on truly independent photons. In the experiment two photons from two 
such singlets interfere at a beam splitter, and as a result the other 
two photons---which nowhere interacted and whose paths nowhere 
crossed---exhibit a 100\%\ correlation in polarization, even when 
no polarization has been measured in the first two photons. The 
propsed experiment permits closure of the remaining loopholes in the 
Bell theorem proof, reveals the quantum nonlocality as a property of 
selection, and pioneers an experimental procedure for exact preparation
of unequal superposition.

\end{abstract}

\bigskip


\section{INTRODUCTION}
\label{INT}
The fourth order interference of photons has been given a growing 
attention in the literature in the last few years mostly because 
it provided several rather unexpected results which 
differ from the classical intensity interference counterparts.
\cite{paul86,ou88,gm87,ohm87,ogmm88,ogmm89,om88,ohm88,gryu88,hom89,om89,%
oum89,rart90,ozwm90,owzm90,cst90,wzm91,wzm92,yurke92,zeil93,p93,p94,ps94}  
Let us mention the down--converted induced 
coherence,\cite{wzm92} non--dependence of the interference 
on the relative intensity of the incoming beams,\cite{oum89}  
a disproval\cite{paul86,owzm90} of Dirac's dictum: \it Interference 
between two different photons never occurs\/\rm,\cite{dir58}  
interference of photons of different colors,\cite{ogmm88} 
entaglement of photons which did not in 
any way directly interact whith each other in the configuration 
space \cite{zeil93} and in the spin space,\cite{p93,ps94} as well as 
particularly successful testing of  
both local \cite{om88,gryu88,rart90}        
and nonlocal \cite{wzm91,wzm92} hidden variable theories.    

In this paper we close the \it no enhancement\/ \rm and 
\it low efficiency\/ \rm loopholes in the Bell theorem proof, 
show that quantum nonlocality is essentially a property of 
selection, and establish a procedure for recording unequal 
superpositions without loss of detection counts.
In accomplishing these objectives we rely on the 
spin features of the interferencethe fourth order 
on a beam splitter which we previously used for 
an entaglement of two photon pairs coming out from 
two cascade sources.\cite{p93,ps94} In the interference 
both, polarized and unpolarized incident photons emerge 
from two different sides of the beam splitter unpolarized but 
correlated.~(Section \ref{sec:BS}) This enables us to devise an 
experiment in which two photons from two such singlet states 
interfere in the fourth order on a third beam splitter and 
as a result two other companion photons from each pair turn out 
entangled and correlated in polarization even when we do not measure 
polarization on the first two at all. In Section \ref{sec:exp} 
we elaborate the theory of such an entaglement and in Section 
\ref{sec:r-exp} we present the experiment in a realistic approach 
discussing the spatial visibility of the correlations.   

Correlated photons come out from cascading atoms in all 
directions allowed by such a three--body process and by registering 
only those pairs which reach detectors we actually select a 
subset of all correlated photons for which one can raise doubts 
as to whether it properly represents the whole set. 
An affirmative assumption, known as the \it no enhancement\/ \rm 
assumption, has been been widely adopted since Clauser and
Horne\cite{cl-h74} first made it. Recently however, 
Santos\cite{san91} surfaced the problem and calculated that no 
experiment carried out on the photons born in a cascade process 
can confirm the assumption. As opposed to this situation photons 
coming from a beam splitter build spin (polarization) correlated 
pairs \it only\/ \rm in particular precisely determined directions 
but, on the other hand, it was believed that such set--ups force 
the experimentalists to discard more than 50\%\ of the data 
because detectors cannot tell one photon from two and one has 
to rely only on coincidence counts. We, however, show that one 
can devise an experiment in which no data need to be discarded 
thus avoiding Santos' objection. We do this in Section 
\ref{sec:r-exp} by providing a device for preselecting spin 
directed correlated photons among those photons which have not in 
any way directly interacted with each other. The experiment can 
close all the existing loopholes in disproofs of local hidden 
variable theories, including the \it low effiency\/ \rm one, 
thanks to preselection of photons and might provide a scheme 
for disproving nonlocal theories as well. In closing the 
\it low effiency\/ \rm lopphole we find out how to measure 
unequal superpositions exactly as presented at the end of 
Section \ref{sec:r-exp}.
 
\section{SPIN INTERFEREOMETRY ON A BEAM SPLITTER}
\label{sec:BS}
 
Our experimental set--up rests on the fact that under particular 
conditions the interference of the fourth order makes unpolarized 
and independent incident photons correlated in polarization 
(spin) and turns polarized incident photons into unpolarized ones.
We recognized this property only recently\cite{p94} because 
although the interference of the fourth order in the 
configuration space has been elaborated in detail in the literature 
\cite{paul86,ou88,gm87,ohm87,ogmm88,om88,ohm88,gryu88,hom89,om89,%
oum89,rart90,ozwm90,owzm90,cst90,wzm91,wzm92} it lacked   
a detailed elaboration and apparently a proper understanding in 
the spin space. One of the rare partial elaborations was  
provided by Ou, Hong, and Mandel for a special case of orthogonally 
polarized photons.\cite{ohm87b} They clearly recognized that 
orthogonally polarized photons incoming to a symmetrically positioned 
beam splitter produce a \it singlet--like\/ \rm state at a beam 
splitter \cite{ou88,om88,ohm88,ohm87b} and that parallelly polarized 
photons incoming to a symmetrically positioned beam splitter never 
appear on its opposite sides \cite{hom87} but it does not seem 
to have been recognized that the polarization of incoming photons 
does not have any effect on the correlation in polarization of the 
outgoing photons and that it only affects the intensity of the 
photons emerging from the opposite sides of the beam splitter. 
In the following we carry out the spin elaboration of the fourth 
order interference on a beam splitter using some results obtained 
by Pavi\v ci\'c.\cite{p94} 

Let two photons interfere on a beam splitter as shown in 
Fig.~\ref{bs}. First, we describe the interference of polarized 
and later on of unpolarized photons. The state of incoming 
polarized photons is given by the product of two prepared 
linear--polarization states: 
\begin{eqnarray}
|\Psi\rangle=\left(\cos\theta_{1_0}|1_x\rangle_{1_0}\>+
\>\,\sin\theta_{1_0}|1_y\rangle_{1_0}\right)
\otimes\left(\cos\theta_{2_0}|1_x\rangle_{2_0}\>+
\>\,\sin\theta_{2_0}|1_y\rangle_{2_0}\right)\,,\label{eq:2-state}
\end{eqnarray}
where $|1_x\rangle$ and $|1_y\rangle$ denote the mutually orthogonal 
photon states. So, e.g., $|1_x\rangle_{1_0}$ means the upper incoming 
photon polarized in  
direction $x$. If the beam spliter were removed it would cause 
a \it ``click''\/ \rm at the detector D1 and no \it ``click''\/ \rm 
at the detector D1$^\perp$ provided the birefringent polarizer P1 is 
oriented along $x$. Here D1$^\perp$ means a detector counting photons 
coming out at the \it other exit\/ \rm of the birefringent prism P1. 
Angles $\theta_{1_0}$,$\theta_{2_0}$ are the angles along which 
incident photons are polarized with respect to a fixed direction. 

We do not consider any interference of the second order because 
the signal and idler photons emerging from the nonlinear 
crystals which we use in our experiment in Sec.~\ref{sec:exp} 
have random phases relative to each other. Thus we are left 
with the interference of the fourth order, i.e., with two 
interacting photons described by two corresponding electric fields. 
To describe the appropriate interaction 
of photons with the beamspliter, polarizers, and detectors we 
make use of the second quantization formalism employed, e.g., 
by Paul,\cite{paul86} Mandel, Ou, Hong, Zou, and Wang, 
\cite{ohm87,ozwm90,owzm90} and Campos, Saleh, and Teich.\cite{cst90}

We introduce polarization by means of the stationary electric field 
operator whose orthogonal components read (see Fig.~\ref{bs})
\begin{equation}
\hat E_j({\bf r}_j,t)=\hat
a_j(\omega_j)e^{i\mbox{\scriptsize\bf k}_j\cdot
\mbox{\scriptsize\bf r}_j-i\omega_jt}\,. 
\end{equation}

The anihilation operators describe joint actions of polarizers, 
beam splitter, and detectors. The operators act on the states as 
follows: ${\hat a}_{1x}|1_{x}\rangle_1=|0_{x}\rangle_1$, \ 
${\hat a}_{1x}^{\dagger}|0_{x}\rangle_1=|1_{x}\rangle_1$, \  
${\hat a}_{1x}|0_{x}\rangle_1=0$, etc. Thus, the action of the 
polarizers P1,P2 and detectors D1,D2 can be expressed as: 
\begin{equation}
{\hat a}_i={\hat a}_{ix\,out}\cos\theta_i+
{\hat a}_{iy\,out}\sin\theta_i\,,\label{eq:D2} 
\end{equation}
where $i=1,2$. 

The operators corresponding to the other choices of detectors we 
obtain accordingly. E.g., the action of the polarizer P2 and the 
corresponding detector D2$^\perp$ (as shown in Fig.~\ref{exp}) 
is described by
\begin{equation}
{\hat a}_2=-{\hat a}_{2x\,out}\sin\theta_2+
{\hat a}_{2y\,out}\cos\theta_2\,.\label{eq:D2'-perp} 
\end{equation}

The outgoing electric--field operators describing photons which pass 
through beam splitter BS and through polarizers P1 and P2 and are 
detected by detectors D1 and D2 will thus read
\FL
\begin{eqnarray}
\hat E_1=\left(\hat a_{1x}t_x\cos\theta_1+\hat
a_{1y}t_y\sin\theta_1\right)
e^{i\mbox{\scriptsize\bf k}_1\cdot
\mbox{\scriptsize\bf r}_1-
i\omega_1t_1}+i\left(\hat a_{2x}r_x\cos\theta_1+\hat
a_{2y}r_y\sin\theta_1\right)
e^{i\tilde{\mbox{\scriptsize\bf k}}_2\cdot
\mbox{\scriptsize\bf r}_1-
i\omega_2t_1}\,,\label{eq:e1}
\end{eqnarray}
\begin{eqnarray}
\hat E_2=\left(\hat a_{2x}t_x\cos\theta_2+\hat
a_{2y}t_y\sin\theta_2\right)e^{i\mbox{\scriptsize\bf k}_2
\cdot\mbox{\scriptsize\bf r}_2-i\omega_2t_2}
+i\left(\hat a_{1x}r_x\cos\theta_2+\hat
a_{1y}r_y\sin\theta_2\right)e^{i\tilde{\mbox{\scriptsize\bf k}}_1
\cdot\mbox{\scriptsize\bf r}_2-i\omega_1t_2}\,,\label{eq:e2}
\end{eqnarray}
where $i$ assures the phase shift during the reflection on the beam 
splitter, $t_j$ is the time of detection of a photon by detector 
D$_j$, $\omega_j$ is the frequency of photon $j$, $c$ is the 
velocity of light. Here the crystal as a supposed source of (idler 
and signal down--converted) photons is assumed to be positioned
symmetrically to the beam splitter (with respect to the photon paths 
from the center of the crystal to the beam splitter). This is just 
the opposite to the elaboration carried in Pavi\v ci\'c\cite{p94} 
where detectors were assumed to be positioned symmetrically to the 
beam splitter while time delays for the sources were introduced in 
order to describe photons born in atomic cascade processes used in   
Pavi\v ci\'c and Summhammer.\cite{ps94} 

The joint interaction of both photons with the beam splitter, 
polarizers P1,P2, and detectors D1,D2 is given by a 
projection of our wave function onto the Fock vacuum space 
by means of $\hat E_1$,$\hat E_2$ wherefrom we get the 
following probability of detecting photons by D1,D2:\cite{p94}
\begin{equation}
P(\theta_{1_0},\theta_{2_0},\theta_1,\theta_2)
=\langle\Psi|\hat E_2^\dagger\hat E_1^\dagger
\hat E_1\hat E_2|\Psi\rangle
=A^2+B^2-2AB\cos\phi\,,\label{eq:prob} 
\end{equation}
where $|\Psi\rangle$ is given by Eq.~(\ref{eq:2-state}) and where 
\begin{equation}
\phi=(\tilde{\mbox{\boldmath$k$}}_2-
\mbox{\boldmath$k$}_1)\cdot\mbox{\boldmath$r$}_1+
(\tilde{\mbox{\boldmath$k$}}_1-
\mbox{\boldmath$k$}_2)\cdot\mbox{\boldmath$r$}_2+
(\omega_1-\omega_2)(t_1-t_2)\,,\label{eq:phi}
\end{equation}
$A=S_{1'1}(t)S_{2'2}(t)$ and $B=S_{1'2}(r)S_{2'1}(r)$, where 
\begin{equation}
S_{ij}=s_x\cos\theta_{i}\cos\theta_{j}
+s_y\sin\theta_{i}\sin\theta_{j}\,.
\end{equation}

Assuming $\omega_1=\omega_2$ we obtain (see Fig.~\ref{bs}) 
$\phi=2\pi(z_2-z_1)/L\,$, where $L$ is the spacing of the 
intereference fringes.\cite{ou88} 

For $t_x=t_y=r_x=r_y=2^{-1/2}$ and $\cos\phi=1$
(we can modify $\phi$ by moving the detectors transversely to the 
incident beams) the probability reads
\begin{eqnarray}
P(\theta_{1_0},\theta_{2_0},\theta_1,\theta_2)=(A-B)^2= 
{1\over4}\sin^2(\theta_{1_0}-\theta_{2_0})\sin^2(\theta_1-\theta_2)\,,
\label{eq:coinc} 
\end{eqnarray}
which for removed polarizers makes
\begin{eqnarray}
P(\theta_{1_0},\theta_{2_0},\infty,\infty)= 
{1\over2}\sin^2(\theta_{1_0}-\theta_{2_0})\,.\label{eq:coincinf} 
\end{eqnarray}

We see that the probability in Eq.~(\ref{eq:coinc}) 
factorizes (see Fig.~\ref{bs}) left--right
(corresponding to 1$_0$--2$_0$--preparation $\leftrightarrow$
D1--D2--detections) and not up--down (corresponding to
${\mbox{\scriptsize\rm 1$_0$}\,\atop\mbox{\scriptsize\rm
2$_0$}}\!\!\updownarrow$ preparation) in spite of the up--down 
initial independence described by the product of the upper and 
lower function in Eq.~(\ref{eq:2-state}). We also see  
that by changing the relative angle between the polarization planes 
of the incoming photons we only change the light intensity of the 
photons emerging from the beam splitter at particular sides. Thus 
the photons either emerge on two different sides of the beam splitter 
correlated according to Eq.~(\ref{eq:coinc}) or both emerge on 
one side according (when we do not measure their outgoing 
polarization) to the following overall probability 
\begin{eqnarray}
P(\theta_{1_0},\theta_{2_0},\infty\times\infty)= 
{1\over2}\bigl[1+\cos^2(\theta_{1_0}-\theta_{2_0})\bigr]\,,
\label{eq:both} 
\end{eqnarray}
which together with Eq.~(\ref{eq:coincinf}) adds up to one.

We also see that the photon beams leave the beam splitter unpolarized:
\begin{eqnarray}
P(\theta_{1_0},\theta_{2_0},\theta_1,\infty)
={1\over4}\sin^2(\theta_{1_0}-\theta_{2_0})\,.
\label{eq:unpol}
\end{eqnarray}

If both incoming photons come in unpolarized --- coming, e.g., 
from two simultaneously cascading independent atoms or better from 
two other beam splitters what as a possibility directly follows from 
just obtained Eq.~(\ref{eq:unpol}) --- they appear\cite{p94} 
correlated whenever they appear at the opposite sides of the 
beam splitter:
\begin{eqnarray}
P(\infty,\infty,\theta_{1},\theta_{2})=
{1\over2}\sin^2(\theta_{1}-\theta_{2})
\label{eq:coinc-inf} 
\end{eqnarray}
and partially correlated whenever they both emerge from one side of 
the beam splitter:  
\begin{eqnarray}
P(\infty,\infty,\theta_{1},\theta_{2})= 
{1\over2}\bigl[1+\cos^2(\theta_{1}-\theta_{2})\bigr]\,.
\label{eq:both2} 
\end{eqnarray}
The latter probability can be checked experimentally with the help 
of an additional beam splitter in each arm following 
Rarity and Tapster\cite{rart89} or by means of photons of different 
colors which one can distinguish using frequency 
filters (prisms).\cite{ogmm88,om88b,diff-c93}  

In case of nondegenerate idler and signal down--converted 
photons (by means of asymmetrically positioned pinholes), 
i.e.~in case of photons of different colors we should, 
according to Eq.~(\ref{eq:phi}), obtain a space--time combination 
of space--like intensity interference and time--like 
frequency--difference beating. The latter effect, however,  
cannot be measured together with observing the intensity 
interference fringes because the fast photon beating would wipe 
out the spatial fringes. For observation of the beating itself 
one uses the optical path--lenght difference method by which the 
coincidences are recorded.\cite{ogmm88,om88b} So, in our notation 
we simply drop the dot products in Eq.~(\ref{eq:phi}) and then 
the method consists in moving the beam splitter up or down in order 
to obtain the optical path--lenght difference $\delta=c|t_1-t_2|$ and 
thus have $|\phi|=|\omega_1-\omega_2|\delta/c$. In this way one can 
register beating corresponding to 30$\>$fs by means of detectors and  
counters whose resolving time is 10$\>$ns.\cite{ogmm88} 
Our main coincidence probability for particular polarization 
measurements given by Eq.~(\ref{eq:prob}) remains the same for the 
beating between photons of different frequencies as it was for the 
degenerate idler and signal photons. The fact that we can trace the 
path of each photon is here not contradictory because, first, we deal 
not with the beam intensity but with the intensity correlation, and, 
secondly, as we already stressed, polarization \it preparation\/ 
\rm of photons is \it erased\/ \rm by the beam splitter anyhow. 

The most important consequence of the obtained equations 
for our experiment is that the photons appear entangled in a singlet 
state whenever they appear on different sides of the beam splitter
provided the condition $\phi=0$ is satisfied no matter whether 
incident photons were polarized or not. For, Eqs.~(\ref{eq:coinc}) 
and (\ref{eq:coinc-inf}) tells us that the probability of \it such\/ 
\rm photons passing parallel polarizers is equal to zero. 

\section{THEORY OF THE ENTAGLEMENT IN THE EXPERIMENT}
\label{sec:exp}

A schematic representation of the experiment is shown in
Fig.~\ref{exp}. Two independent beam splitters BS1 and BS2 
act as two independent sources of two independent singlet pairs 
which is enabled by Eq.~(\ref{eq:coinc}) as elaborated in 
the previous section. Two photons from each pair interfere 
on the beam splitter BS and as a result the other two
photons, under particular conditions elaborated below, 
appear to be in the singlet state although the latter photons 
are completely independent and nowhere interacted. 

An ultrashort\cite{u-s84} laser beam (a subpicosecond one) of 
frequency $\omega_0$ simultaneously (split by a beam splitter) 
pumps up two nonlinear crystals NL1 and NL2 producing in each of 
them pairs of signal and idler photons (simulataneously and  
with equal probability) of frequencies $\omega_1$ and $\omega_2$, 
respectively, which satisfy the following energy and momentum 
conservation conditions: $\omega_0=\omega_1+\omega_2$ and 
$\mbox{\boldmath$k$}_0=\mbox{\boldmath$k$}_1+
\mbox{\boldmath$k$}_2$.\cite{ghom86} By means of the appropriately 
symmetrically positioned pinholes we select half--frequency 
sidebands so as to have $\omega_2=\omega_1$. The idler and signal 
photon pairs coming out from the crystals do not have definite 
phases\cite{hom87,owzm90b} with respect to each other and 
consequently one can have a second order interference neither on 
BS1 nor on BS2. In order to prevent any coherence which might be 
induced by the split pumping beam between the idler (or signal) 
photon from the first crystal and the idler (or signal) photon from 
the second crystal we introduce a phase modulator (which rotates to 
and fro at random and destroy the second order phase coherence) 
following Ou, Gage, Magill, and Mandel.\cite{ogmm89} (We do take a 
correction term corresponding to the modulator into account when 
estimating the visibility below, but do not show it in the equations 
for the sake of their simplicity.) 

Thus, two \it sources\/ \rm BS1 and BS2 both simultaneously emit two 
photons in the singlet states given by Eq.~(\ref{eq:coinc}) to the 
left and to the right. But before we put beam splitters BS1 and BS2 
in place we first have to adjust detectors the beam splitter and 
D1--D2$^\perp$ so as to obtain $\phi=0$. After that we take out BS 
and put in BS1 and BS2 to adjust them and detectors D1'--D2'$^\perp$ 
(while leaving detectors D1--D2$^\perp$ fixed) so as to obtain pure 
singlet states coming out from BS1 and BS2. It follows from 
Eq.~(\ref{eq:coinc}) and Fig.~\ref{exp} that we can do this for 
$\phi=0$ by reaching the minimum of coincidences (ideally the 
minimum should be zero) for $\theta_{1'}=\theta_1$ for BS1 and 
for $\theta_{2'}=\theta_2$ for BS2. It is interesting that this 
step of tuning BS1,BS2 and D1'--D2'$^\perp$ is not crucial, 
because the four photon entanglement is not 
dependent on the positions of D1'--D2'$^\perp$ detectors in 
directions perpendicular to the photon paths, i.e., according to 
Eq.~(\ref{eq:prob-4}) there are no interference fringes for 
photons 1' and 2' --- only for photons 1 and 2. Then we put beam 
splitter BS in place and four photons form elementary quadruples 
of counts which add up to the below calculated probabilities in the 
long run. The quadruple recording is obtained by the following \it 
preselection\/ \rm procedure: whenever exactly two of the \it 
preselection detectors\/ \rm D1--D2$^\perp$ fire in coincidence 
(see Fig.~\ref{exp-0}) a gate for counters D1'--D2'$^\perp$ 
opens. In case only one or none of the so preselected 
D1'--D2'$^\perp$ detectors fires we discard the records (because 
they correspond to four or three photons detected by 
D1--D2$^\perp$, respectively). In case exactly two of four 
D1'--D2'$^\perp$ detectors fire, the corresponding counts 
contribute to our statistics. The possibility of two photons 
going into one arm of the beam splitter as well as the 
possibility that a detector fails to react because of 
its inefficiency we discuss in Sec.~\ref{sec:r-exp}.  

The state of the four photons immediately after leaving BS1 and BS2 
from their oppposite sides is described by the product of the two 
superpositions corresponding to singlet pairs produced --- 
according to Eq.~(\ref{eq:coinc}) --- on BS1 and BS2, respectively: 
\begin{eqnarray}
|\Psi\rangle={1\over\sqrt2}\bigl(|1_x\rangle_{1'}|1_y\rangle_1\>-
\>\,|1_y\rangle_{1'}|1_x\rangle_1\bigr)
\otimes{1\over\sqrt2}\bigl(|1_x\rangle_{2'}|1_y\rangle_2\>-
\>\,|1_y\rangle_{2'}|1_x\rangle_2\bigr)\,,\label{eq:4-state}
\end{eqnarray}
where $|1_x\rangle$ and $|1_y\rangle$ denote the mutually orthogonal 
photon states. 

The annihilation of photons at detectors D1',D2' after passing the 
polarizers P1',P2' (oriented at angles $\theta_{1'},\theta_{2'}$) 
are described by the following electric field operators 
\begin{eqnarray}
\hat E_{1'}=
(\hat a_{1'x}\cos\theta_{1'}+\hat a_{1'y}\sin\theta_{1'})
e^{-i\omega_1't_{1'}}\,,\label{eq:D1'}
\end{eqnarray}
\begin{eqnarray}
\hat E_{2'}=
(\hat a_{2'x}\cos\theta_{2'}+\hat a_{2'y}\sin\theta_{2'})
e^{-i\omega_2't_{2'}}\,.\label{eq:D2'}
\end{eqnarray}
Here, phases of the photons which accumulate between beam splitters 
BS1,BS2 and detectors D1',D2' add the factors 
$e^{-i\omega_jt_j}$, where $t_j$ is the time of detection of a 
photon by detector D$_j\!$' and $\omega_j$ is the frequency of the 
photon. [Until Eq.~(\ref{eq:prob-4}) we shall consider the 
frequencies of photons different for the sake of generality.] 
   
The electric outgoing field operators describing photons which pass 
through beam splitter BS, polarizers P1,P2 and detectors D1,D2 are 
given by Eqs.~(\ref{eq:e1}) and ({\ref{eq:e2}). 

The joint interaction of all four photons with the beam splitter, 
polarizers P1--P2', and detectors D1--D2'$^\perp$ is given by the 
following projection of our initial state given by 
Eq.~(\ref{eq:4-state}) wave function onto the Fock vacuum space: 
\begin{eqnarray}
\hat E_{1'}\hat E_{2'}\hat E_1\hat E_2|\Psi\rangle=
{1\over2}(A\varepsilon_{12}-B\tilde\varepsilon_{12})\varepsilon
|0\rangle\,,\label{eq:fock}
\end{eqnarray}
where $|\Psi\rangle$ is given by Eq.~(\ref{eq:4-state}), 
where $\varepsilon_{12}=\exp\bigl[i\left(
\mbox{\boldmath$k$}_1\cdot\mbox{\boldmath$r$}_1+
\mbox{\boldmath$k$}_2\cdot\mbox{\boldmath$r$}_2-
\omega_1t_1-\omega_2t_2\right)\bigr]$, 
$\tilde \varepsilon_{12}=\exp\left[i\left(\tilde{\mbox{\boldmath$k$}}_1
\cdot\mbox{\boldmath$r$}_2+
\tilde{\mbox{\boldmath$k$}}_2\cdot\mbox{\boldmath$r$}_1-
\omega_1t_2-\omega_2t_1\right)\right]$,  
$\varepsilon=\exp\Bigl[-i\left(\omega_1't_{1'}+
\omega_2't_{2'}\right)\Bigr]$, 
and $A=Q(t)_{1'1}Q(t)_{2'2}$ and $B=Q(r)_{1'2}Q(r)_{2'1}$, where 
\begin{eqnarray}
Q(q)_{ij}=q_x\sin\theta_i\cos\theta_j-q_y\cos\theta_i\sin\theta_j\,. 
\end{eqnarray}

The corresponding probability of detecting all four photons by
detectors D1--D2'$^\perp$ is thus
\begin{eqnarray}
P(\theta_{1'},\theta_{2'},\theta_1,\theta_2)
=\langle\Psi|\hat E_{2'}^\dagger\hat E_{1'}^\dagger\hat E_2^
\dagger\hat E_1^\dagger
\hat E_1^{\phantom\dagger}\hat E_2^{\phantom\dagger}
\hat E_{1'}^{\phantom\dagger}\hat E_{2'}^{\phantom\dagger}|\Psi\rangle
={1\over4}(A^2+B^2-2AB\cos\phi)\,,\label{eq:prob-4} 
\end{eqnarray} 
where 
\begin{equation}
\phi=(\tilde{\mbox{\bf k}}_2-
\mbox{\bf k}_1)\cdot\mbox{\bf r}_1+
(\tilde{\mbox{\bf k}}_1-
\mbox{\bf k}_2)\cdot\mbox{\bf r}_2+(\omega_1-\omega_2)(t_1-t_2)\,. 
\end{equation}
For $\omega_1=\omega_2=\omega_1'=\omega_2'$ we obtain (see 
Fig.~\ref{bs} which applies on BS from Fig.~\ref{exp} as well) 
$\phi=2\pi(z_2-z_1)/L\,$, where $L$ is the spacing of the 
intereference fringes. $\phi$ can be changed by moving the 
detectors transversely to the incident beams. 

To make Eq.~(\ref{eq:prob-4}) more transparent, without loss of 
generality, we here consider 50:50 beam splitter: 
$t_x=t_y=r_x=r_y=2^{-1/2}$. In Sec.~\ref{sec:r-exp} we also 
consider a polarized beam splitter. 

For $\phi=0$ the above probability reads
\begin{eqnarray}
P(\theta_{1'},\theta_{2'},\theta_1,\theta_2)
={1\over4}(A-B)^2=
{1\over16}\sin^2(\theta_{1'}-\theta_{2'})\sin^2(\theta_1-\theta_2)\,.
\label{eq:l-r} 
\end{eqnarray}
We again see that the probability factorizes left--right 
(corresponding to D1'--D2' $\leftrightarrow$ D1--D2 detections --- 
see Fig.~\ref{exp}) and not up--down (corresponding to
${\mbox{\scriptsize\rm BS1}\,\atop\mbox{\scriptsize\rm
BS2}}\!\!\updownarrow$ preparation) as one would be tempted to 
conjecture from the product of the upper and lower function 
in Eq.~(\ref{eq:4-state}). For removed polarizers P1,P2 
Eq.~(\ref{eq:l-r}) gives:
\begin{eqnarray}
P(\theta_{1'},\theta_{2'},\infty,\infty)
={1\over8}\sin^2(\theta_{1'}-\theta_{2'})\,.\label{eq:bell}
\end{eqnarray}

The overall probability of detecting both photons in one arm of 
BS is given by: 
\begin{eqnarray}
P(\theta_{1'},\theta_{2'},\theta_1\times\theta_2)
={1\over16}\bigl[\cos(\theta_{1'}-\theta_1)
\cos(\theta_{2'}-\theta_2)+\cos(\theta_{1'}-\theta_2)
\cos(\theta_{2'}-\theta_1)\bigr]^2.\protect\label{eq:two} 
\end{eqnarray}
which for removed polarizers reads: 
\begin{eqnarray}
P(\theta_{1'},\theta_{2'},\infty\times\infty)
={1\over8}[1+\cos^2(\theta_{1'}-\theta_{2'})]\,.\label{eq:bellx2}
\end{eqnarray}
The latter probability one obtains so as to add up all 
the probabilities of detecting polarizations of each photon in one 
arm, i.e., $P(\theta_{1'},\theta_{2'},\theta_1\times\theta_2)$ 
[given by Eq.~(\ref{eq:two})], 
$P(\theta_{1'},\theta_{2'},\theta_1\times\theta_2^\perp)$, etc. 
We see that the probabilities (\ref{eq:bell}) and (\ref{eq:bellx2}) 
add up to one. 

The probability (\ref{eq:l-r}) shows that for $\phi=0$ by removing 
one of the polarizers we lose any left--right (Bell--like) spin 
correlation completely: $P(\theta_{1'},\infty,\theta_1,\theta_2)
={1\over16}\sin^2(\theta_1-\theta_2)$. On the other hand, for 
$\phi\ne0$ we obtain a partial left--right correlation even when 
two polarizers, one on each side, are removed.

\section{THE EXPERIMENT AND THE BELL ISSUE}
\label{sec:r-exp}

The main point of our experiment is that the correlation 
between photons 1' and 2', i.e., between photons which never 
interacted in the past, persists even when we do not measure 
polarization on their companions 1 and 2 at all as follows 
from Eq.~(\ref{eq:bell}). Therefore we shall concentrate on the 
experiment without polarizers P1,P2 behind beam splitter BS. 
To make our point we present the appropriate experimental set--up 
in a simplified and reduced scheme presented in Fig.~\ref{exp-0}. 
The set--up deals with four photons of the same frequency and 
relies on (computer) time windows for coincidence detections 
which compensate for the long responding time of the detectors. 
Afterwards we shall consider the experiment in a more realistic 
approach making use of polarizers P1,P2 as shown in Fig.~\ref{exp}.  

In the idealized approach from Sec.~\ref{sec:exp}. 
the probability of detecting all four photons by D1--D2'$^\perp$ 
in coincidence for 50:50 beam splitter for $\phi=0$ and with equal 
time delays (that is for a completely symmetrical position of BS) 
is given by Eq.~(\ref{eq:bell}) and the probability of detecting 
both photons in one of the arms by Eq.~(\ref{eq:bellx2}). We see 
that these two probabilitites add up to 1/4. (The other 3/4 
correspond to \it orthogonal detections\/ \rm by D$^\perp$ 
detectors.) The former probability given by Eq.~(\ref{eq:bell}) 
and describing coincidence detections by D1' and D2' corresponds 
--- when multiplied by 4 --- to the following singlet state:
\begin{eqnarray}
|\Psi_s\rangle={1\over\sqrt{2}}(|1_x\rangle_1|1_y\rangle_2-
|1_y\rangle_1\>|1_x\rangle_2)\,.\label{eq:s-state}
\end{eqnarray}
Multiplication by 4 is for photons which emerge from the same 
side of BS and which therefore do not belong to our statistics. 
Analogously, the probability of coincidental detection by 
D1' and D2'$^\perp$ (which we will make use of later on), given by   
Sec.~\ref{sec:r-exp})  
\begin{eqnarray}
P(\theta_{1'},\theta_{2'}^\perp,\infty,\infty)
={1\over8}\cos^2(\theta_{1'}-\theta_{2'})\,.\label{eq:triplet}
\end{eqnarray}
corresponds to the following triplet--like state:
\begin{eqnarray}
|\Psi_t\rangle={1\over\sqrt{2}}(|1_x\rangle_1|1_y\rangle_2+
|1_y\rangle_1\>|1_x\rangle_2)\,.\label{eq:t-state}
\end{eqnarray}

Thus, photons 1',2' belonging to quadruples containing 
photons 1,2, which appear at different sides of 
the beam splitter behave \it quantum--like\/ \rm showing 
--- according to Eq.~(\ref{eq:bell}) --- 100\%\ \it relative 
modulation\/\rm.\cite{om88} In other words, by detecting the 
right photons on different sides of the beam splitter we \it 
preselect\/ \rm  the orthogonal individual left photons pairs 
(25\%\ of all pairs) with probability one, while by detecting 
the right photons both on one side of the beam splitter we 
(\it would\/ \rm have --- if it had been experimentally possible) 
preselect the parallel pairs (75\%\ of all pairs) with 
probability 1/3. When we compare this result with the classical 
formulation of Pavi\v ci\'c\cite{p94} carried out by Paul and 
Wegmann\cite{paul94} we see that the former case (photons 
emerging from different sides of the beam splitter) is 
``completely non--classical.'' This means that it is the 
nonclassical feature of the intensity correlations that enables 
our experiment. 

Let us now dwell on the details of the experiment without polarizers 
P1,P2 behind beam splitter BS as shown in Fig.~\ref{exp-0}. A pair 
consisting of two photons 1' and 1 appears from BS1 simultaneously 
with another pair 2'--2 on BS2. Photons 1', 2', and 1 \it and\/ 
\rm 2 are directed towards detectors D1' \it or\/ \rm D1'$^\perp$, 
D2' \it or\/ \rm D2'$^\perp$, and D1 \it and\/ \rm D2, respectively. 
Of all detections registered by D1,D2 only those counts which
occur within a short enough time windows (about 10ns) are fed
to the \it preselection coincidence counter\/\rm. Thanks to 
the ultrashort pumping beam ($\omega_0$), which ensure an average 
appearance of down--converted pairs of photons ($\omega=\omega_0/2$ 
--- coming out from the crystals and passing through symmetrically 
positioned pinholes) every 50$\>$ns, we are able to effectively 
control coincidences each of which occurs (as a property of 
downconversion) well within our time windows. In this way we 
overcome the problem of having the detector reaction time longer 
than the fourth order correlation time and the coherence time. 
So, each pair of the pulses belongs to the two photons which 
interfered on BS so as to appear at the opposite sides of the beam 
splitter. (Realistically, as we will see below, it boils down 
to about 85\%; A possibility of having detected 3 or 4 photons 
due to a possibility of both photons emerging from one side of 
BS1 or BS2 we resolve below.) Each D1--D2 time window 
is coupled (as calculated from the time--of--flight 
difference) with a computer gate for counts from detectors 
D1',D2',D1'$^\perp$,D2'$^\perp$. If D1,D2 
counters do not register coincidence counts but a only a single 
count, then the ``gated'' D1',D2',D1'$^\perp$,D2'$^\perp$ 
recordings are discarded. If they do, we get potential data for 
our statistics what we call \it Bell recording\/ \rm in 
Fig.~\ref{exp-0}. Since we use birefringent polarizers we 
have to have a coincidence firing of exactly two of the counters 
D1',D2',D1'$^\perp$,D2'$^\perp$ in order to obtain definite data 
for the statistics. Firing of one or none of the counters as well 
as of three or all four discard the corresponding data because 
they do not belong to our set of quadruple events. 
$P(\theta_{1'},\theta_{2'},\infty\times\infty)$ of 
Eq.~(\ref{eq:bell}) is then given by the following ratio between 
the numbers of coincidence counts: 
\begin{eqnarray}
f(\theta_{1'},\theta_{2'})={n(\rm{D1'}\cap\rm{D2'})\over 
n[(\rm{D1'}\cup\rm{D1'}^\perp)\cap(\rm{D2'}\cup
\rm{D2'}^\perp)]}\,.\label{eq:freq}
\end{eqnarray}
divided by 4. Division by 4 compensates for the photons which 
emerged from the same side of BS and were therefore discarded 
from the statistics as not belonging to the considered set of 
events. Of course, we produce an error here because counters 
can remain inactive because of their inefficiency but we can 
always make use of Mach--Zehnder interferometers instead of 
BS1 and BS2 to avoid this problem. Their adventages would be, 
first, that we can adjust them so that photons almost always 
emerge from the oposite sides of their second beam splitters 
and almost never from the same sides and, secondly, that a 
detector resolution time which is much longer then the coherence 
time is not any more a problem (in contradistinction to a 
single beam splitter) --- it is even required.\cite{ozwm90,cst90} 
We did not use the interferometers here so as not to 
overcomplicate our presentation, but we will comment on them 
in some detail later on. Alternatively we can use photons of 
different frequencies for each pair and relying on their 
beating instead on the spatial fringes as explained at the 
end of Sec.~\ref{sec:BS}.   

The assumed 100\%\ visibility above is of course an 
oversimplification since the measurement of probability 
(\ref{eq:prob-4}) cannot be measured at a point 
(see Fig.~\ref{bs}) but only over a detector width $\Delta z$. 
Therefore, in order to obtain a more realistic probability 
following Ghosh and Mandel\cite{gm87} we integrate 
Eq.~(\ref{eq:prob-4}) over $z_1$ and $z_2$ over $\Delta z$ 
to obtain 
\begin{eqnarray}
{\cal P}(\theta_{1'},\theta_{2'},\theta_1,\theta_2)
=&&{1\over4}\int_{z_1-\Delta z/2}^{z_1+\Delta z/2}
\int_{z_2-\Delta z/2}^{z_2+\Delta z/2}
\Bigl[A^2+B^2-2AB\cos\bigl[2\pi(z_2-z_1)/L\bigr]\Bigr]
dz_1dz_2\nonumber\\
=&&{1\over4}(A^2+B^2-v2AB\cos\phi)\,,\label{eq:nu}
\end{eqnarray} 
where $v=\bigl[\sin(\pi\Delta z/L)/(\pi\Delta z/L)\bigr]^2$ 
is the \it visibility\/ \rm of the coincidence counting. 
Visibility of 95\%\ has been estimated as achiveable in 
principle,\cite{oum89} 80\%\ and 87\%\ has been reached 
recently.\cite{hi-ef91,hi-ef93} 

Thus, Eq.~(\ref{eq:bell}) corrected for a realistic visibility 
reads:
\begin{eqnarray}
P(\theta_{1'},\theta_{2'},\infty,\infty)
={1\over8}[1-v\cos^2(\theta_{1'}-\theta_{2'})]\,.\label{eq:vis} 
\end{eqnarray} 

To see that our results really tighten all the remaining loopholes 
in disproving local hidden varable theories, let us in the 
end discuss the corresponding Bell's inequality: 
\begin{eqnarray}
S\equiv P(\theta_{1'},\theta_{2'})-P(\theta_{1'},\theta_{2'}')+
P(\theta_{1'}',\theta_{2'}')+P(\theta_{1'}',\theta_{2'})-
P(\theta_{1'}',\infty)-P(\infty,\theta_{2'})\leq0
\,,\label{eq:bell-in} 
\end{eqnarray} 
where $P(\theta_{1'},\theta_{2'})=
4P(\theta_{1'},\theta_{2'},\infty,\infty)$, etc. 
The singlet states of photons 1' and 2' and the corresponding 
probabilities ${1\over2}\sin^2(\theta_{1'}-\theta_{2'})$ 
correspond to the D1'--D2' coincidence counts preselected by D1--D2 
coincidence counts. Since, ideally, no one of so preselected 
photons escapes detection we have thus satisfied Santos' 
demand.\cite{san91} To be more specific, 
$P(\theta_{1'},\theta_{2'})$ is not obtained as 
a coincidence counting rate like in the previous 
experiments\cite{ou88,om88,ohm88} but as the ratio (frequency) 
$f(\theta_{1'},\theta_{2'}$) given by Eq.~(\ref{eq:freq}) where 
the total number of counts in the denominator can actually be 
recorded. 

Ideally, for a violation of Bell's 
inequality, and hence for a possible exclusion of hidden variable 
theories, $v$ must be\cite{ohm88} larger than $2^{-1/2}$. 
If we also take into account the overall efficiency of detectors 
$\eta$ defined by 
$P(\theta_{1'},\theta_{2'})=\eta f(\theta_{1'},\theta_{2'})$
for the case of equal superposition given by Eq.~(\ref{eq:4-state}) 
the inequality (\ref{eq:bell-in}) can be violated only 
if\cite{san91a,gam87} 
\begin{eqnarray}
\eta(1+v\sqrt2)>2\,.\label{eq:eff}
\end{eqnarray}
So, for the visibility $v=1$ we must have $\eta>83\%$. 
For the recently achieved visibilities $v=0.8$\cite{hi-ef93} and 
$v=0.87$\cite{hi-ef91} according to Eq.~(\ref{eq:eff}) this means 
$\eta>0.94$ and $\eta>0.9$ which is already anounced as 
achievable.\cite{hi-ef93a,ci-e94} So, the experiment in the presented 
set--up is just about to be feasible. However, using our most 
recent result we can adjust it so as to be comfortably 
over this verge and conclusively feasible with the present
technology. Let us elaborate this in some detail.  

As ``forerunners'' of our singlet states \it selected\/ \rm 
among photons whose paths nowhere crossed in the space, several 
simpler set--ups involving only two fotons interfering on beam 
splitters were reported. In particular, Mach--Zehnder 
interferometer was recognized as a possible source 
of 100\%\ correlated photons (i.e., without both photons 
emerging from the same sides of the second beam 
splitter).\cite{ozwm90,cst90} Only, until our result\cite{p94} 
it was not recognized that these photons appear correlated in
polarization and automatically satisfy Santos' demand up to the 
efficiency of detectors. But, after Kwiat, Eberhard, Steinberg, 
and Chiao\cite{ci-e94} in the meantime carried out an 
explicit calculation for the single Mach--Zehnder interferometer 
they immediatelly addressed detector efficiency limitations and 
focussed a recent result by Eberhard\cite{eber93} as a possible 
remedy. (It should be stressed here that in the light of dector 
efficiencies Hardy's\cite{har93} proposal cannot be considered 
as an answer to Santos' objection because today's visibility in 
his proposal is 30\%.) Eberhard has shown that if one used 
unequal superpositions 
\begin{eqnarray}
|\Psi_r\rangle={1\over\sqrt{1+r^2}}(|1_x\rangle_1|1_y\rangle_2+
r|1_y\rangle_1\>|1_x\rangle_2)\label{eq:r-state}
\end{eqnarray}
instead of equal ones (given by $r$=1), then one would be able 
to lower the required efficiency of detectors down to 67\%. 
The problem was how to achieve this. Eberhard himself 
connected the effect with the background noise and the 
drawback of this definition was, first, that one can hardly 
specify \it the background\/ \rm and, secondly, that one loses 
counts. We have however found the following way how to use 
Eberhard's result without any losses and without invoking any 
background noise. 

From Eqs.~(\ref{eq:prob-4}) and (\ref{eq:D2'-perp}) it follows 
that the probability of having coincidence counts by detectors 
D1' and D2'$^\perp$ after a selection by (see Fig.~\ref{exp}) 
detectors D1 and D2 with the orientation of polarizers 
$\theta_2=0$ and $\theta_1=\pi/2$ and with  $t_y=r_y=2^{-1/2}$ 
is given by
\begin{equation}
P(\theta_{1'},\theta_{2'}^\perp)={1\over1+r^2}
(\cos\theta_{1'}\cos\theta_{2'}+
r\sin\theta_{1'}\sin\theta_{2'})^2\,,\label{eq:r-prob}
\end{equation}
where $r={r_x\over t_x}$ and where we also take 
counts registered by D1'$^\perp$ and D2' into account 
in order to obtain the proper probability.  Since it can easily be 
shown that the detected photons are in the state described by 
Eq.~(\ref{eq:r-state}) we have thus recognized Eberhard's term $r$ 
as the ratio between the reflection and transmission coefficient 
of the polarized beam splitter. So, for $r=0.31$, 
i.e., for $T_x=t_x^2=0.91$, an efficiency greater than 
70\%\ suffices for a loophole free Bell's experiment depending 
on the visibility on the beam splitter. On the other hand, 
Eq.~(\ref{eq:r-prob}) establishes an experimental procedure 
for measuring unequal superposition without loss of detection 
counts since the probability $P(\theta_{1'},\theta_{2'}^\perp)$ can 
be obtained as the frequency 
\begin{eqnarray}
f(\theta_{1'},\theta_{2'}^\perp)={n(\rm{D1'}\cap\rm{D2'}^\perp)
\over n[(\rm{D1'}\cup\rm{D1'}^\perp)\cap(\rm{D2'}\cup
\rm{D2'}^\perp)]}\,.\label{eq:freq-r}
\end{eqnarray}
where both $n$(D1'$\cap$D2'$^\perp$) and 
$n$[(D1'$\cup$D1'$^\perp$)$\cap
$(D2'$\cup$D2'$^\perp$)] can be recorded with equal accuracy. 

This hopefully\cite{p90} closes all the remaining loophoes in 
the Bell's proof, and constitutes a most discriminating test of 
Bell's inequality. 

\section{CONCLUSION}
\label{sec:dis}

The experiment we proposed is a realization of a polarization 
correlation between two independent and unpolarized photons.  
The experiment is based on a newly discovered nonclassical 
effect in the interference of the fourth order on a beam splitter 
according to which two unpolarized incident photons emerge 
from a beam splitter correlated in polarization as follows 
from Eqs.~(\ref{eq:prob}) and (\ref{eq:coinc-inf}). 
The essential new element of the experiment is that it 
puts together two photons from two singlets formed on two beam 
splitters, makes them interfere on a third beam splitter, and 
as a result we find polarization correlations between 
the other two photons which nowhere interacted and whose paths 
nowhere crossed even when no polarization measurement have been 
carried out on the former two photons as follows from 
Eqs.~(\ref{eq:prob-4}) and (\ref{eq:l-r}). As for the latter 
two photons which nowhere interacted, one of their subsets 
turns out to contain only photons in the singlet state and since 
we are able to extract these photons with probability one 
one can consider them \it preselected\/ \rm by their 
pair--companion photons which interfered on the beamsplitter. 
By using birefringent prisms we can in principle detect all 
the photons from the subset and obtain the probability 
(\ref{eq:bell}) as a proper frequency (a ratio of counts) 
given by Eq.~(\ref{eq:freq}). In this way we close the \it 
enhancement loop\/ \rm of the Bell theorem proof. On the other 
hand, the experiment shows that it is not a direct interaction 
between photons or their common origin what entagles them 
in a polarization singlet state but particular correlations which 
one can preselect without resorting to polarization measurement 
at all. We conclude that nonlocality is essentially a property of 
selection. This might exclude all nonlocal hidden--variable  
theories which rely on some kind of a physical entaglement 
via a common origin.   

The realistic estimation of the experiment carried out in 
Sec.~\ref{sec:r-exp} for the equal superposition given by 
Eq.~(\ref{eq:s-state}) shows that such a set--up is just about to 
be feasible within the so--called (see Sec.~\ref{sec:r-exp}) \it 
83\%\ limit\/ \rm thus narrowing the second, \it efficiency\/ 
\rm loophole in the Bell theorem proof. It should be 
stressed here that a very helpful feature of the 
considered effect is that the entaglement (quadruple 
firing of \it preselection detectors\/ \rm D1--D2 behind the 
beam splitter \it and\/ \rm two of D1'--D2$^\perp$ catching 
the other two ``free'' photons) is independent of positions of 
D1'--D2$^\perp$ and also of the moment of their firing as follows 
from Eqs.~(\ref{eq:prob-4}) and (\ref{eq:nu}). In other words 
the visibility of the whole entaglement and the visibility 
of the two photon coincidence on BS practically do not differ. 

To narrow down the efficiency loophole completely we resort to 
the polarization measurement and unequal superposition [given by 
Eq.~(\ref{eq:r-state})] whereby we can make the experiment 
comfortably within the so--called \it 67\%\ limit\/ \rm by 
recognizing Eberhard's \it r--term\/ \rm not as a measure of 
a background noise but as the ratio of the reflection to the 
transmission coefficient in one of the measured polarization 
directions (on a polarized BS). At the same time 
this approach establishes to our knowledge the first 
experimental procedure for \it exact\/ \rm measurement of 
unequal superposition without loss of detection 
counts. In other words, when we, with the help of partially 
polarized beam splitter BS and detectors D1--D2$^\perp$, 
preselect photons in the subset of photons in the unequal 
triplet--like state given by Eq.~(\ref{eq:r-state}) we do not 
lose counts because the detectors by means of the 
birefringent polarizers P1' and P2' register all counts so 
that we can form a proper frequency, given by Eq.~(\ref{eq:freq-r}), 
in order to verify the corresponding probability, given by 
Eq.~(\ref{eq:r-prob}). This closes the \it efficiency\/ \rm loop 
in the Bell theorem proof. 

\acknowledgments
The author is grateful to his hosts Prof.~K.--E.~Hellwig, 
Institut f\"ur Theoretische Physik, TU Berlin 
and Dr.~J.~Summhammer, Atominstitut der \"Osterreichischen 
Universit\"aten, Vienna for their hospitality and to them and 
Prof.~H.~Paul, Humboldt--Universit\"at zu Berlin for valuable 
discussions. He also acknowledges supports of the Alexander von 
Humboldt Foundation, the Technical University of 
Vienna, and the Ministry of Science of Croatia.

\begin{figure}
\caption{Beam splitter.}
\label{bs} 
\end{figure}

\begin{figure}
\caption{Outline of the experiment.}
\label{exp} 
\end{figure}

\begin{figure}
\caption{Reduced scheme of the experiment.}
\label{exp-0} 
\end{figure}

\vfill\eject 

\vfill 

\vbox to 4cm{\vfill}

\epsffile{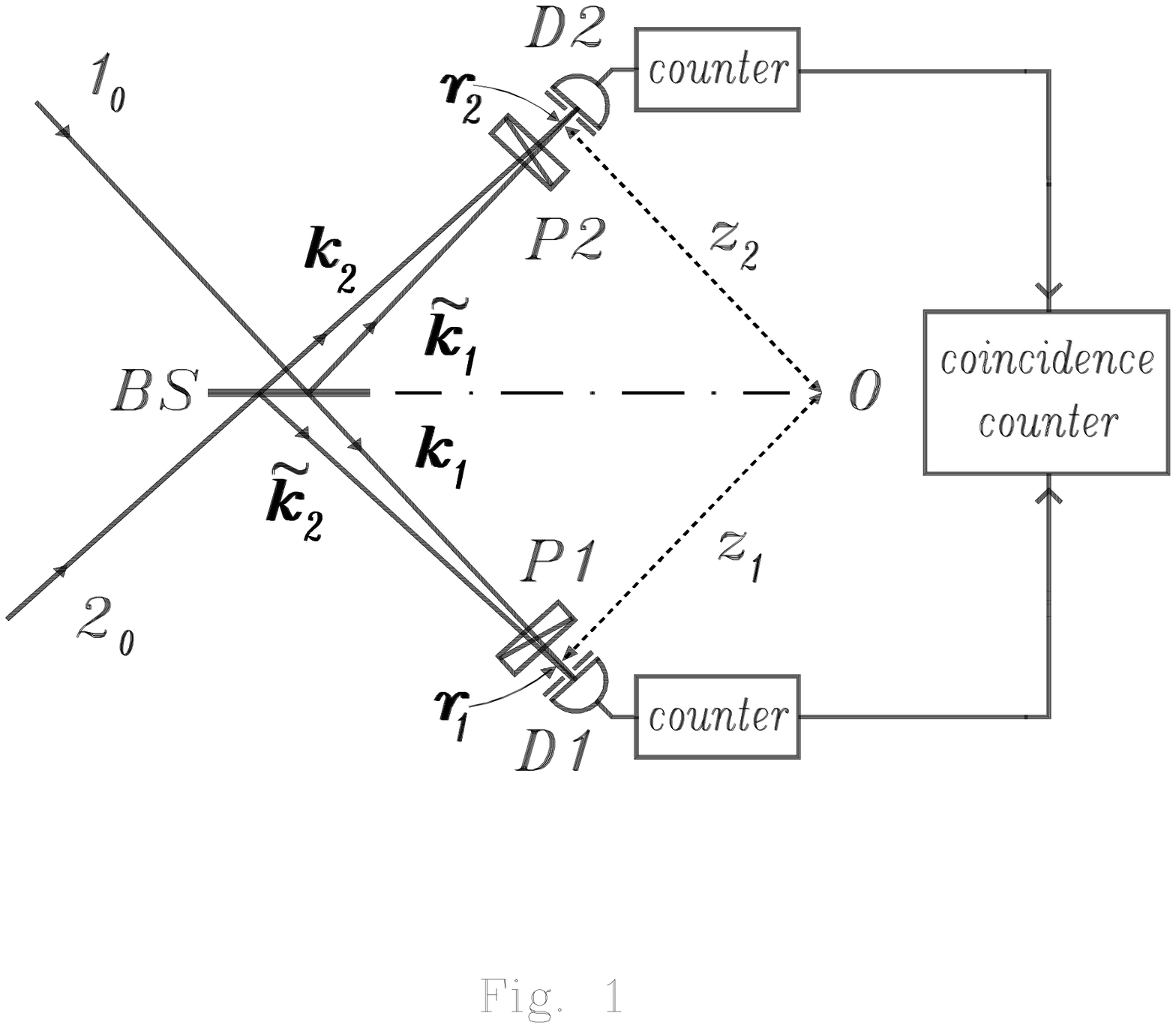}

\vfill\eject 

\vfill

\vbox to 1cm{\vfill}

\epsffile{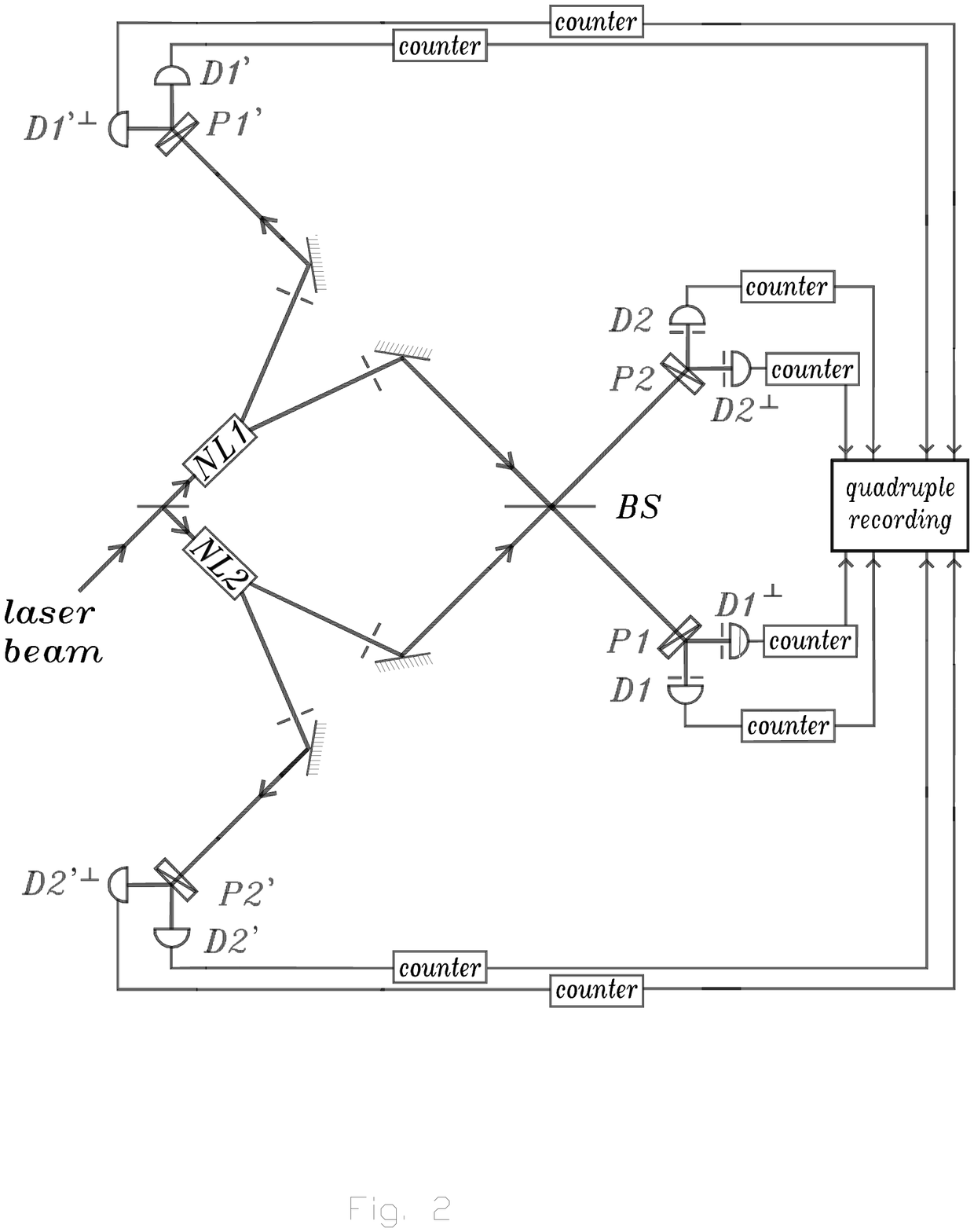}

\vfill\eject 

\vfill

\vbox to 1cm{\vfill}

\epsffile{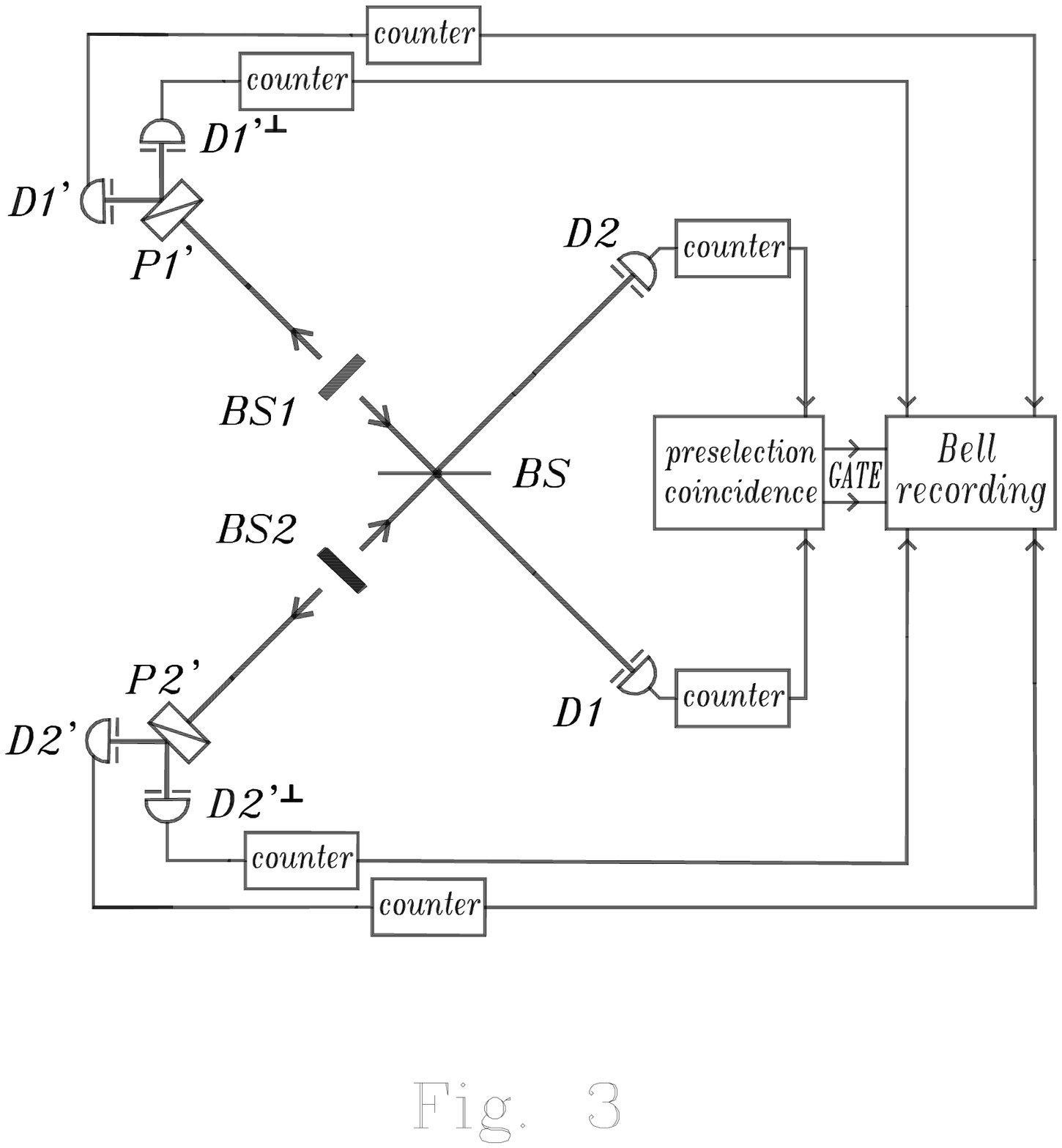}

\end{document}